\documentstyle[11pt,newpasp,twoside,psfig]{article}
\def\edcomment#1{\iffalse\marginpar{\raggedright\sl#1\/}\else\relax\fi}
\reversemarginpar

\def\spose#1{\hbox to 0pt{#1\hss}}
\def\lta{\mathrel{\spose{\lower 3pt\hbox{$\mathchar"218$}}
     \raise 2.0pt\hbox{$\mathchar"13C$}}}
\def\gta{\mathrel{\spose{\lower 3pt\hbox{$\mathchar"218$}}
     \raise 2.0pt\hbox{$\mathchar"13E$}}}

\def\=#1{\overline{#1}}

\def\deg{^\circ}             

\def\kpc{{\rm\,kpc}}

\def\msun{{\rm\,M_\odot}}
\def\lsun{{\rm\,L_\odot}}
\def\yr{{\rm\,yr}}

\def\gyr{{\rm\,Gyr}}

\def\etal{{et~al.\ }}
\def\aa #1 #2 {A\&A, #1, #2}
\def\aas #1 #2 {A\&AS, #1, #2}
\def\acm #1 #2 {ACM-Trans Math Software, #1, #2}
\def\ada #1 #2 {Ann Astrophys, #1, #2}
\def\agabstr #1 #2 {Astr Ges Abstr Ser, #1, #2}
\def\aj #1 #2 {AJ, #1, #2}
\def\anach #1 #2 {Astr Nachr, #1, #2}
\def\apj #1 #2 {ApJ, #1, #2}
\def\apjl #1 #2 {ApJL, #1, #2}
\def\apjs #1 #2 {ApJS, #1, #2}
\def\araa #1 #2 {ARAA, #1, #2}
\def\apss #1 #2 {ApSpaceS, #1, #2}
\def\celmech #1 #2 {Cel Mech, #1, #2}
\def\esom #1 #2 {ESO Messenger, #1, #2}
\def\fundcp #1 #2 {FunCosP, #1, #2}
\def\jcp #1 #2 {J Comp Phys, #1, #2}
\def\jfm #1 #2 {J Fluid Mech, #1, #2}
\def\jmp #1 #2 {J Math Phys, #1, #2}
\def\ma #1 #2 {Mitt Astr Ges, #1, #2}
\def\mn #1 #2 {MNRAS, #1, #2}
\def\nat #1 #2 {Nat, #1, #2}
\def\obs #1 #2 {Observatory, #1, #2}
\def\pasj #1 #2 {PASJ, #1, #2}
\def\pasp #1 #2 {PASP, #1, #2}
\def\phyr #1 #2 {PhysRep, #1, #2}
\def\physd #1 #2 {Physica D, #1, #2}
\def\rpp #1 #2 {RepProgPhys, #1, #2}
\def\ssr #1 #2 {Sp Sci Rev, #1, #2}

\begin{document}
\title{Formation of the Galactic Halo}
 \author{Ortwin Gerhard}
\affil{Astronomisches Institut, Universit\"at Basel\\ Venusstrasse 7,
CH-4102 Binningen, Switzerland, \\ email: Ortwin.Gerhard@unibas.ch}

\begin{abstract}
  Recent observational and theoretical work suggests that the
  formation of the Galactic stellar halo involved both dissipative
  processes and the accretion of subfragments. With present data, the
  fraction of the halo for which an accretion origin can be
  substantiated is small, of order 10 percent.  The kinematics of the
  best halo field star samples show evidence for both dissipative and
  dissipationless formation processes.  Models of star-forming
  dissipative collapse, in a cosmological context and including
  feedback from star formation, do not confirm the simple relations
  between metallicity, rotation velocity, and orbital eccentricity for
  halo stars as originally predicted. The new model predictions are
  much closer to the observed distributions, which have generally been
  interpreted as evidence for an accretion origin. These results are
  broadly consistent with a hierarchical galaxy formation model, but
  the details remain to be worked out.
\end{abstract}

\section{Introduction}

In our quest to understand the formation of galaxies, the Milky Way
takes a special place. Nowhere else can the properties of stellar
populations be observed in comparable detail, can ages, metallicities,
and kinematics be determined for as many individual stars. The
Galactic stellar halo, the topic of this article, gives a vivid
illustration: with its mass of only some $10^9\msun$ and estimated
surface brightness near the Sun of 27.7 V mag/arcsec$^2$ (Morrison
1993) it would be hard to observe at all in external galaxies. Yet it
is believed to be the oldest component of the Galaxy, and to hold
important information about the Milky Way's formation process.

The early debate and even much of the recent discussion about the
formation of the Galactic stellar halo focussed on two contrasting
scenarios. One view, based on an apparent correlation between the
metallicities and the orbital velocities and eccentricities of halo
stars, held that the halo formed in a rapid dissipative collapse phase
during which metal enrichment took place (Eggen, Lynden-Bell \&
Sandage 1962, ELS).  The other, based on the lack of abundance
gradient in the halo globular clusters and a $\sim 10^9\yr$ age spread
inferred from their horizontal branch colours, proposed that that the
Galactic halo formed by the prolonged, chaotic accretion of dwarf
galaxy-like fragments (Searle \& Zinn 1978).  Because in modern
samples of low-metallicity halo stars no correlation is in fact found
between metallicity and kinematics (Norris 1986, Carney, Latham \&
Laird 1990, Chiba \& Beers 2000), and because of direct evidence for
accretion such as in the form of the Sgr dwarf galaxy (Ibata, Gilmore
\& Irwin 1994), the accretion scenario has become the standard view in
the field (Freeman \& Bland-Hawthorne 2002).

However, recent observational and theoretical developments suggest
that some revision of this picture may be necessary.  The fraction of
the halo for which an accretion origin can be substantiated with
present data is only of order 10 percent (Section 2).  Modern analysis
shows no evidence of an age spread for the metal-poor globular
clusters within an error of $\sim 10^9\yr$; these are consistent with
being old and coeval (Rosenberg \etal 1999, Salaris \& Weiss 2002).
Only intermediate metallicity clusters around [Fe/H]$\sim -1.2$ show
evidence for an age spread of $2-3\gyr$.  From the new large samples
of halos stars there is evidence for both dissipative and
dissipationless processes during halo formation (Section 3). Recent
models of dissipative collapse, in a cosmological context and
including feedback from star formation, show that the predictions of
the ELS model on which much of this discussion is based are
oversimplified and partly incorrect (Section 4). And finally, in the
current hierarchical models for structure formation in the
Universe, accretion may occur both in a smooth, dissipative form and
through the merging of subunits. In the hierarchical framework, the
interesting question regarding the origin of the halo is not
``dissipative collapse {\sl or} merging?'', but
\begin{itemize}
\item how important was smooth accretion compared to lumpy accretion
or merging?
\item did small units form stars before they fell together?
\item were infalling subunits tidally disrupted in the halo or
did they survive into the disk or center? 
\end{itemize}
Understanding the formation of the Galactic halo from the observed
properties of its constituents will help in answering these
questions within the more general problem of galaxy formation
in hierarchical models of structure formation in the Universe.

\section{Evidence for accretion}

The most dramatic case for accretion in the Galactic halo is provided
by the disrupting Sagittarius dwarf galaxy discovered by Ibata, Gilmore
\& Irwin 
(1994). The Sgr dwarf is visible over some $20\deg\times 8\deg$ on the
sky, centered at (l,b)$\simeq(6\deg,-14\deg)$, at a galactocentric
distance of $16\kpc$ on the other side of the Galactic center. It is
orientated roughly perpendicular to the Galactic plane along its
orbit. Sgr contains $L_V\simeq1-2\times 10^7\lsun$ of stars, 4
globular clusters, and probably $\sim 10^9\msun$ of dark matter, which
are in the process of being added to the Galactic halo (Ibata \etal
1997). The extended stream of previously dissolved stars has been
found in several surveys.

With time, such streamers will phase-mix and become invisible in
photometric surveys unless confined to particular orbital planes.
However, substructure in phase-space is preserved much longer, once
the Galactic potential has settled and evolves only slowly. Helmi
\etal (1999) identified one sub-group of halo stars in a Hipparcos
sample which they identified as the remains of an ancient accretion
event. This subgroup was confirmed by Chiba \& Beers (2000) in their
larger sample of halo stars. According to these authors, the total
mass of the precursor object is a few percent of the present Galactic
halo, i.e, similar to that of the Sgr dwarf.  New surveys based on
halo Carbon stars (Ibata \etal 2001), RR Lyrae stars (Vivas \etal
2001), SDSS (Newberg \etal 2002), and GDSS (Kundu \etal 2002) have
found at most a very small number of streams other than the Sgr
stream. At least half the halo Carbon stars appear to belong to the
Sgr stream according to Ibata \etal (2001), severely limiting the
amount of stars that can have been accreted in the last 5 Gyr.

Thus from present data, the fraction of the Galactic halo for which
there is evidence for accretion is of order 10\%. This is in some
contrast to the prediction of Bullock, Kravtsov \& Weinberg (2001),
who find that in hierarchical CDM models a large number of tidal
streams from disrupted dwarf galaxies should make up a large fraction
of the Galactic stellar halo. One way out would be that many of the
small dark matter lumps predicted in CDM never made stars (see, e.g.,
Bullock, Kravtsov \& Weinberg 2001, Stoehr \etal 2002). Alternatively,
it is possible that the planned GAIA satellite will in fact detect
halo substructure for a much larger fraction of halo stars, but on
much finer scales. This would be the signature expected from accretion
events during the early formation of the Galaxy, now in a more
advanced state of phase-mixing, or from accretion of much smaller
subunits. However, in this case the interpretation must involve
questions about star formation issues as well. What are the smallest
and what are the typical star-forming units in a Galactic collapse? In
the Galactic disk, there is clearly structure on the scale of giant
molecular clouds up to $\sim 10^6\msun$.  And how can we distinguish
between an inevitably lumpy and prolonged collapse, and the accretion
of separate fragments?

\section{Evidence from halo star properties}

There is general agreement that the Galactic halo is the oldest
component of the Milky Way. The most metal-poor globular clusters with
[Fe/H]$<-1.5$ are old and coeval, with age $12\pm1 \gyr$ (Rosenberg
\etal 1999, Salaris \& Weiss 2002).  Intermediate metallicity clusters
around [Fe/H]$\sim -1.2$ show evidence for an age spread of $2-3\gyr$
and a slightly younger age.  Halo field stars with [Fe/H]$<-1.8$ have
red turnoff colours B-V$\simeq 0.4$, and comparison with isochrones
likewise indicates old ages. Only a small fraction of stars at higher
metallicities have bluer turnoff colours, limiting the number of dwarf
galaxies with intermediate age populations that could have been
accreted (Unavane, Wyse \& Gilmore 1996).

A detailed analysis of 1200 stars with [Fe/H]$\le -0.6$, distance
estimates, radial velocities, and proper motions was published by
Chiba \& Beers (2000).  From this large, kinematically unbiassed
sample they deduced some noteworthy properties of Galactic halo stars:
There is no correlation between [Fe/H] and orbital eccentricity $e$
for the metal-poor stars, demonstrating that the evidence given by
Eggen \etal (1962) is a result of kinematic bias.  However, there is a
concentration of halo stars on radial orbits at [Fe/H]$\simeq -1.7$,
which may be a signature of early collapse, and a concentration of
disk stars on near-circular orbits at [Fe/H]$>-1$.  Other than this,
the [Fe/H]-eccentricity diagram is populated remarkably uniformly;
even at the lowest abundances, $\sim 20\%$ of the stars have $e<0.4$.
Furthermore, while confirming the absence of a correlation of rotation
velocity with [Fe/H] for [Fe/H]$\le -1.5$, Chiba \& Beers (2000) find
a decrease of rotation velocity with vertical height even for the
lowest metallicity halo stars. This they interpret as a signature of
dissipative formation of the inner halo. The inferred density
distribution of the halo is nearly spherical in the outer region
beyond $R=15-20\kpc$, and highly flattened in the inner region.
Studying the halo star orbits while slowly removing the Galactic disk
potential, Chiba \& Beers (2001) concluded that the inner halo prior
to disk formation must have been substantially rounder than now, with
axis ratio $c/a\simeq 0.8$, but still more flattened than the outer
halo.

Based on their results, Chiba \& Beers (2000) argue for a hybrid
picture, in which the inner halo formed by dissipative contraction,
while the outer halo was made mainly by the accretion of subgalactic
fragments.  (They discuss previous related ideas.) This would also be
consistent with the tangentially anisotropic velocity distribution of
BHB field stars at large distances inferred by Sommer-Larsen \etal
(1997). However, because no division of the inner and outer halo
components is visible in either the density profile or the rotation
properties, this suggests that both dissipative and dissipationless
processes may have occurred simultaneously at each radius, with the
dissipationless processes relatively more important at larger radii.
As discussed below, this would fit in quite naturally with the
evolution expected in clumpy collapses in CDM models.

\section{Modern dissipative collapse models}

In modern computer models, it is possible to simulate the collapse,
dissipation, star formation, and enrichment in an assembling galaxy in
considerable detail, including the effects of a two-phase medium and
of the feedback from supernovae.  To compare with the abundances and
kinematics of halo stars, one must follow the enrichment and
kinematics of successive stellar generations.  In such a simulation,
it is not yet possible to follow the entire evolution from the initial
gravitational clustering in the large-scale CDM universe to the late
high-resolution baryonic processes.  A number of simplifying
assumptions must therefore be made, as is also the case in, e.g., the
well-known semi-analytic models.  In particular, in the absence of a
quantitative theory of star formation, simple recipes for star
formation rates must be used which, however, can be calibrated against
observed star formation rates such as those of Kennicutt (1998).

In a recent model, Samland \& Gerhard (2003) followed the dynamical
collapse and star formation of a massive disk galaxy within a growing
$\Lambda$CDM dark matter halo, whose mass evolves according to the
cosmological simulations of the VIRGO-GIF project (Kauffmann \etal
1999).  Small-scale structure and merging in the halo were ignored;
this also by-passes the so-called angular momentum problem. The
baryonic matter falls in with the dark matter. Both have the same
angular momentum distribution, which corresponds to $\lambda=0.05$
and is similar to the universal distribution found by Bullock \etal
(2001). The model includes two interstellar medium phases, one a hot
gas fluid, the other a cold/warm cloud medium. These phases and the
stars interact through a number of processes, including the energy
release and enrichment from young stars and supernovae. Stars form
from the cold phase with a rate approximately $\propto \rho^{3/2}$,
and are subsequently followed by an N-body code. Due to the
macroscopic description of these processes, the model contains several
parameters which are not well-determined theoretically. However, most
of these can be calibrated against observations and, because of the
self-regulating nature of the interactions, the dependence of the
physical variables on these parameters is only modest.

The disk galaxy that forms in this model has about three times the
mass of the Galaxy and is not a special model of the Milky Way.
However, several results emerged which are of relevance for the
formation of the Galactic halo. (1) The feedback from supernovae is
important especially at early times when the potential of the dark
matter halo is still shallow. Thus, the collapse and formation
of the most metal-poor component ([Fe/H]$<-1.9$) takes about $1 \gyr$,
significantly longer than their dynamical time. (2) As a result, there
is no dependence of rotation velocity on metallicity for these extreme
halo stars; see Fig.~1. This relation only emerges at higher
metallicities, at [Fe/H]$\simeq -1.8$, similar as for the observed
halo stars. (3) Using the distribution in Fig.~1, it is possible to
identify a number of stellar subcomponents, which can tentatively be
identified with observed components in the Milky Way (see Samland \&
Gerhard 2003 for details).  This is surprising, since no corresponding
information was put into the simulation. (4) Because of the interplay
between enrichment, feedback and dynamical collapse, the distribution
of stars in the [Fe/H] - orbital eccentricity plane is surprisingly
broad (Fig.~2). There does not exist a well-defined relation between
these quantities for the model stars.  The only significant difference
to the observed distribution of Chiba \& Beers (2000) is a lack of
stars with metallicities [Fe/H]$\simeq -2$ on near-circular orbits.

Thus collapse models including cosmological infall and feedback
from supernovae are considerably more complicated and can in some
aspects be qualitatively different from traditional collapse models.
Result (1) is clearly of relevance as regards the distribution of halo
globular cluster ages. Results (2) and (4) are contrary to the
predictions that are usually associated with the Eggen, Lynden-Bell \&
Sandage model. 

Bekki \& Chiba (2001) have published a one-phase model which in terms
of the gas dynamics, star formation, and chemical enrichment
description is much simpler than the model just described, and has no
feedback, but which includes small-scale structure by imprinting a CDM
power spectrum on the initial spherical matter distribution. In their
model, substructure clumps form, and dissipationless processes such as
dynamical friction and merging resp.\ disruption of these clumps are
important. This has the effect of filling in the gap in the
metallicity - eccentricity diagram of metal-poor halo stars on
circular orbits (see their Fig.~13).

Based on these dynamical models, the idea of a hybrid formation for
the Galactic halo looks promising. Within the broad class of
hierarchical models, the dissipative collapse of a distributed gas
component and the accretion of substructure is expected to occur
simultaneously. From the results discussed above, it appears possible
that the right mixture of the two will in the end be able to explain
the observed properties of the Galactic stellar halo.  There is
clearly still much work to do in finding out whether this assertion is
true, and if so, what the right mixture is. By doing this work we will
gain valuable insight into the galaxy formation process at large.

\begin{figure}[t]
\centerline{\vbox{
\psfig{figure=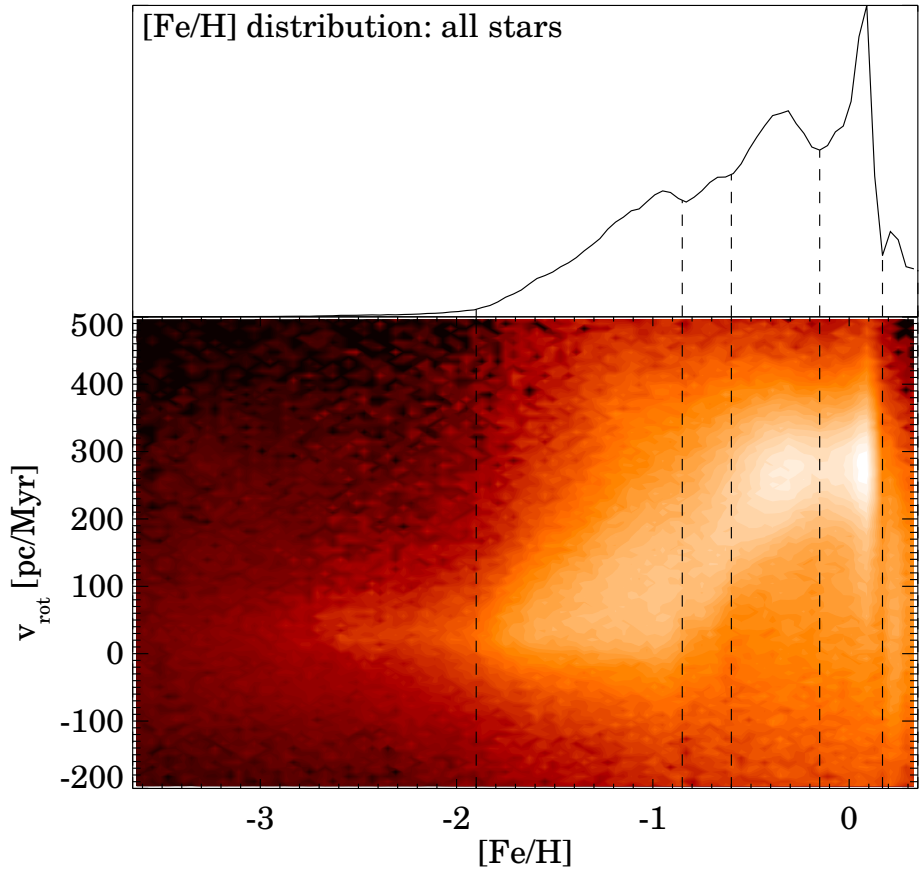,width=\hsize}}}
\caption{Lower panel: distribution of stars in the metallicity-rotation
velocity plane in the model of Samland \& Gerhard (2003). Rotation velocities
are perpendicular to the total angular momentum vector. Upper panel: 
metallicity distribution of all model stars, projecting along the rotation
velocity axis. The dashed lines separate plausibe subpopulations based
on this diagram. }
\end{figure}

\begin{figure}[t]
\centerline{\vbox{
\psfig{figure=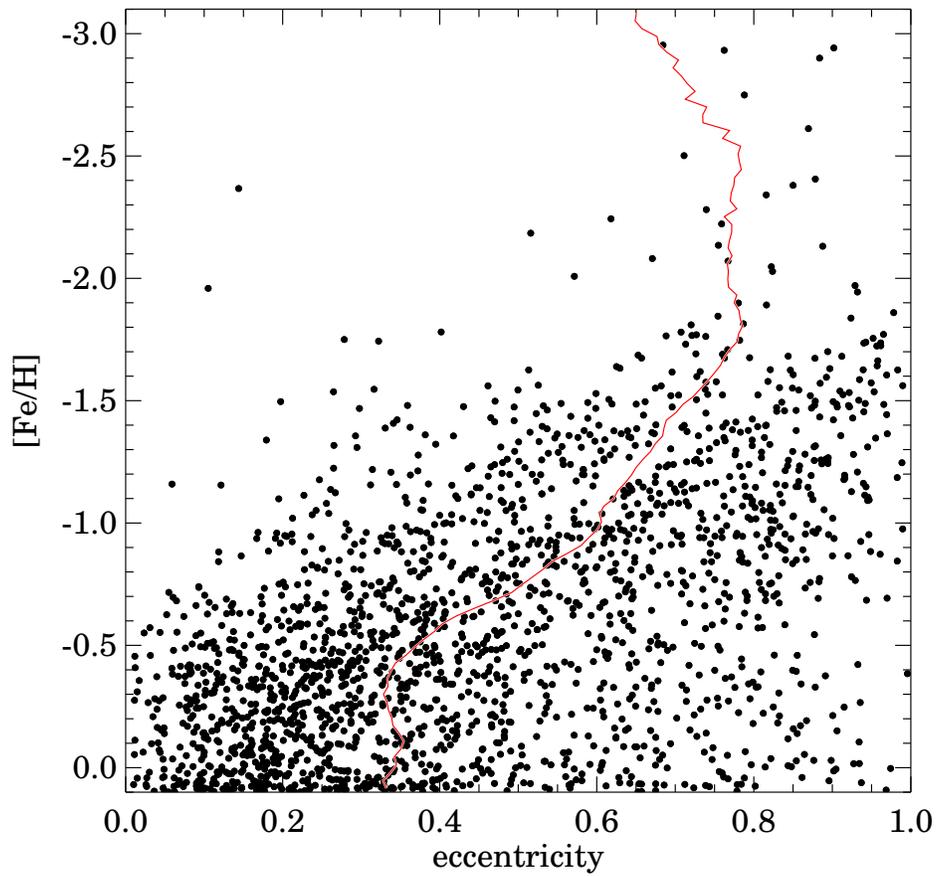,width=\hsize}}}
\caption{Stellar orbital eccentricities for a random sample of all stars
in the model of Samland \& Gerhard (2003), as a function of metallicity.
The line shows the mean eccentricity as function of [Fe/H]. }
\end{figure}

\end{document}